\pdfoutput=1
\documentclass[prx,twocolumn,showpacs,amsmath,amssymb,floatfix,superscriptaddress]{revtex4-1}

\usepackage{graphicx}%Include figure files
\usepackage{dcolumn}%Align table columns on decimal
\usepackage{bm}% bold math$V_q$
\usepackage{hyperref}

\begin{document}

\title{Spin and charge pumping by steady or pulse current-driven magnetic domain wall: A self-consistent multiscale time-dependent-quantum/time-dependent-classical approach} 
		
\author{Marko D. Petrovi\'{c}}
\affiliation{Department of Mathematical Sciences, University of Delaware, Newark,  DE 19716, USA}
\author{Bogdan S. Popescu}
\affiliation{Department of Physics and Astronomy, University of Delaware, Newark, DE 19716, USA}
\altaffiliation{Present Address: Catalan Institute of Nanoscience and Nanotechnology (ICN2), Campus UAB, Bellaterra, 08193 Barcelona, Spain}
\author{Utkarsh Bajpai}
\affiliation{Department of Physics and Astronomy, University of Delaware, Newark, DE 19716, USA}
\author{Petr Plech\'a\v{c}}
\affiliation{Department of Mathematical Sciences, University of Delaware, Newark,  DE 19716, USA}
\author{Branislav K. Nikoli\'{c}}
\email{bnikolic@udel.edu}
\affiliation{Department of Physics and Astronomy, University of Delaware, Newark, DE 19716, USA}
		
\begin{abstract}
We introduce a multiscale framework which combines time-dependent nonequilibrium Green function (TD-NEGF) algorithms, scaling linearly in the number of time steps and describing quantum-mechanically conduction electrons in the presence of time-dependent fields of {\em arbitrary} strength or frequency, with classical time evolution of localized magnetic moments described by the Landau-Lifshitz-Gilbert (LLG) equation. The TD-NEGF+LLG framework can be applied to a variety of problems where  current-driven spin torque induces dynamics of magnetic moments as the key resource for next generation spintronics. Previous approaches to such nonequilibrium many-body system (like steady-state-NEGF+LLG framework) neglect noncommutativity of a quantum Hamiltonian of conduction electrons at different times and, therefore, the impact of time-dependent magnetic moments on electrons leading to pumping of spin and charge currents. The pumped currents can, in turn, self-consistently affect the dynamics of magnetic moments themselves. Using magnetic domain wall (DW) as an example, we predict that its motion will pump time-dependent spin and charge currents (on the top of unpolarized DC charge current injected through normal metal leads to drive the DW motion), where the latter can be viewed as a realization of quantum charge pumping due to time-dependence of the Hamiltonian and left-right symmetry breaking of the two-terminal device structure. The conversion of AC components of spin current, whose amplitude increases (decreases) as the DW approaches (distances from) the normal metal lead, into AC voltage via the inverse spin Hall effect offers a tool to precisely track the DW position along magnetic nanowire. We also quantify the DW transient inertial displacement due to its acceleration and deceleration by pulse current and the entailed spin and charge pumping. Finally, TD-NEGF+LLG as a nonperturbative (i.e., numerically exact) framework allows us to establish the limits of validity of the so-called spin-motive force (SMF) theory for pumped charge current by time-dependent magnetic textures---the perturbative analytical formula of SMF theory becomes {\em inapplicable} for {\em large frequencies} (but unrealistic in magnetic system) and, more importantly, for {\em increasing noncollinearity} when the angles between neighboring magnetic moments exceed $\simeq 10^\circ$.
\end{abstract}

\maketitle

The current-driven dynamics of collinear, such as macrospin~\cite{Ralph2008,Berkov2008,Xiao2005}, and noncollinear, such as domain walls (DWs)~\cite{Tatara2008,Yamaguchi2004,Kim2017a,Lee2004a,Baumgartner2017} and skyrmions~\cite{Nagaosa2013,Fert2013}, textures of localized magnetic moments are both a fundamental problem for nonequilibrium quantum many-body physics and a key resource for next generation spintronics~\cite{Locatelli2014,Kent2015,Grollier2016,Borders2017}. For example, the  current-driven spin torque induced magnetization dynamics in  magnetic tunnel junctions (MTJs)~\cite{Ralph2008,Berkov2008} or ferromagnet/spin-orbit-coupled-material bilayers~\cite{Baumgartner2017,Manchon2018,Nikolic2018}  can implement variety of functionalities, such as nonvolatile magnetic random access memories (MRAM), microwave oscillators, microwave detectors, spin-wave emitters, memristors and artificial neural networks~\cite{Locatelli2014,Kent2015,Grollier2016,Borders2017}. The spin torque can also move DWs and skyrmions along magnetic nanowires which underlies racetrack~\cite{Parkin2008,Parkin2015} and skyrmionic memories~\cite{Koshibae2015}, respectively, with potentially ultralow energy consumption.

\begin{figure}
	\includegraphics[scale=1.0,angle=0]{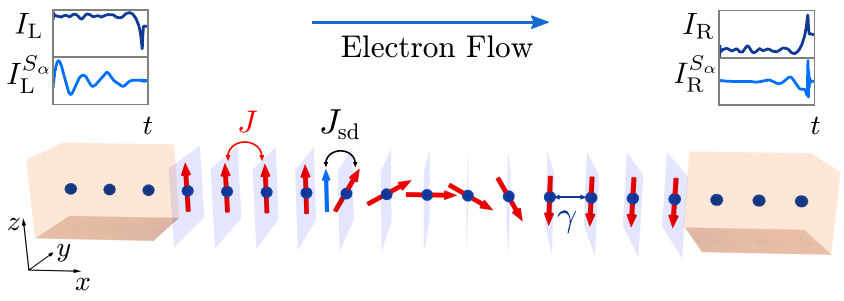}
	\caption{Schematic view of a magnetic nanowire, hosting a DW formed by noncollinear arrangement of localized magnetic moments (red arrows), which is attached to two normal metal leads. The DW dynamics is induced by injecting unpolarized charge current from the NM leads, so that electrons become spin-polarized as they traverse collinear magnetic moments and exert spin torque on those localized  magnetic moments which are noncollinear to their spin polarization vector  (blue arrow). In turn, electrons propagating through time-dependent potential landscape created by the localized magnetic moments pump time-dependent charge  $I_\mathrm{L,R}(t)$ and spin $I_\mathrm{L,R}^{S_\alpha}(t)$ currents into the leads, which are superimposed on charge and spin currents due to the applied DC or pulse bias voltage. The electronic subsystems is modeled on a tight-binding lattice described by the quantum Hamiltonian in Eq.~\eqref{eq:tbh}, whereas the localized magnetic moments are described by the classical Hamiltonian in Eq.~\eqref{eq:heisenberg}.}
	\label{fig:fig1}
\end{figure}

The theoretical analysis of these phenomena requires to account for the interaction of fast conduction electrons, described quantum-mechanically, with slow magnetic moments whose dynamics can be captured by the classical Landau-Lifshitz-Gilbert (LLG) equation~\cite{Berkov2008,Stiles2007,Evans2014}. However,  quantum transport  studies of spin torque in spin-valves~\cite{Haney2007,Wang2008b,Yu2017,Nikolic2018}, MTJs~\cite{Theodonis2006,Heiliger2008,Stamenova2017} and DWs~\cite{Zhang2004a,Garate2009,Balaz2012,Waintal2004,Xiao2006a,Tatara2007,Yuan2016a} are typically confined to computing torque of steady current of electrons acting on a chosen static configuration of localized magnetic moments. Similarly, standard simulations of current-driven magnetization dynamics~\cite{Xiao2005,Berkov2008} or motion of DWs~\cite{Baumgartner2017,Lee2004a,Stiles2007,Li2004,Li2004a,Thiaville2005,Thiaville2007,Martinez2009,Boone2010,Chureemart2011} and skyrmions~\cite{Iwasaki2013a,Sampaio2013}  evade explicit modeling of the flow of conduction electrons by using only classical micromagnetics into which one has to introduce phenomenological terms to describe the so-called adiabatic (when propagating electron spins remain mostly aligned or antialigned with the localized magnetic moments) and nonadiabatic (which can have local~\cite{Zhang2004a,Garate2009,Balaz2012} and nonlocal~\cite{Waintal2004,Xiao2006a,Tatara2007} contributions) spin torques due to flowing electrons. Deriving additional torque expressions is required in the presence of spin-orbit coupling~\cite{Knoester2014,Hals2014} or nontrivial topology~\cite{Braun2012} of magnetic textures~\cite{Akosa2017}. 

A handful of studies~\cite{Ohe2006a,Ohe2006b,Salahuddin2006,Xie2017,Ellis2017} have also attempted to develop a multiscale combination of  computational quantum  (or even simpler semiclassical~\cite{Akosa2015,Claudio-Gonzalez2012,Lee2013a,Sturma2016}) transport of conduction electrons  with discretized  LLG equation for the motion of localized magnetic 
moments described by the classical vectors $\mathbf{M}_i(t)$. However, these attempts employ steady-state nonequilibrium density matrix, strictly applicable only to systems which do not evolve in time, which can be expressed in terms of the lesser Green function $\mathbf{G}^<(E)$ of the nonequilibrium Green function (NEGF) formalism~\cite{Stefanucci2013} 
\begin{equation}\label{eq:steadydm}
{\bm \rho}_\mathrm{neq} = \frac{1}{2 \pi i} \int\limits_{-\infty}^{+\infty} dE\, \mathbf{G}^<(E).
\end{equation}
Thus, such NEGF+LLG approach~\cite{Ohe2006a,Ohe2006b,Salahuddin2006,Xie2017,Ellis2017} naively assumes that electrons respond instantaneously to time-dependent potential introduced into the  quantum Hamiltonian of the conduction electrons by the time evolution of $\mathbf{M}_i(t)$, thereby {\em neglecting  noncommutativity} of the quantum Hamiltonian at different times. On the other hand, it is well-known that even infinitely slow dynamics of $\mathbf{M}_i(t)$ can pump spin currents~\cite{Tserkovnyak2005,Chen2009}, as well as  charge current if additional conditions are satisfied~\cite{Chen2009,Hals2010,Mahfouzi2012}. Therefore, using NEGF+LLG approach precludes taking into account self-consistent feedback~\cite{Lee2013a,Kim2012b} where the dynamics of $\mathbf{M}_i(t)$ leads to pumped spin currents which, in turn, can exert additional torque and time-retarded damping (with microscopically~\cite{Sayad2015,Hammar2016}  rather than phenomenologically~\cite{Bose2011,Thonig2015} determined memory kernel) on  $\mathbf{M}_i(t)$ thereby modifying its dynamics. Finally, time-dependent quantum treatment of electrons is required to describe  pulse-current-induced dynamics of $\mathbf{M}_i(t)$ which is of paramount importance in basic research experiments~\cite{Baumgartner2017} and, e.g., racetrack memory applications~\cite{Parkin2008,Parkin2015} where usage of current pulses~\cite{Thomas2006} or their trains~\cite{Thomas2007} reduces threshold current density to move the DW while precise control of the DW position can be achieved by tailoring pulse duration and shape~\cite{Thiaville2007,Thomas2010,Chauleau2010,Taniguchi2015,Pivano2017}.

Taking into account these effects demands to construct the {\em time-dependent nonequilibrium density matrix}, ${\bm \rho}_\mathrm{neq}(t)$, which can be accomplished using the time-dependent NEGF (TD-NEGF) framework~\cite{Stefanucci2013,Gaury2014}
\begin{equation}\label{eq:timedm}
{\bm \rho}_\mathrm{neq}(t) = \frac{1}{i}\mathbf{G}^<(t,t') |_{t=t'},
\end{equation}
where the lesser Green function $\mathbf{G}^<(t,t')$ depends on two-times in arbitrary nonequilibrium situations~\cite{Stefanucci2013} [in steady-state nonequilibrium it depends on $t-t'$, so it can be Fourier transformed to energy, as utilized in Eq.~\eqref{eq:steadydm}]. Within such more general framework, NEGF+LLG approach corresponds to taking just the lowest order of ${\bm \rho}_\mathrm{neq}(t)$ expanded in power series in small $d\mathbf{M}_i/dt$~\cite{Mahfouzi2016,Bode2012},  so  that $\mathbf{G}(E)$ and $\mathbf{G}^<(E)$ are assumed to depend only parametrically on time and are effectively computed for the frozen-in-time configuration of $\mathbf{M}_i(t)$. For instance, the neglected first order correction contains information about the Gilbert damping term in the LLG equation~\cite{Mahfouzi2016,Bode2012}. 

The nonequilibrium density matrix  yields expectation value of any physical quantity, such as the current-driven (CD) part of nonequilibrium spin density
\begin{eqnarray}\label{eq:spin}
\mathbf{S}_\mathrm{CD}^i(t) & = & \mathbf{S}_\mathrm{neq}^i(t) - \mathbf{S}_\mathrm{eq}^i \nonumber \\
\mbox{} & = & \frac{\hbar}{2} \mathrm{Tr}_\mathrm{spin}\,[{\bm \rho}_\mathrm{neq}(t) {\bm \sigma}] - \frac{\hbar}{2}  \mathrm{Tr}_\mathrm{spin}\,[{\bm \rho}_\mathrm{eq} {\bm \sigma}].
\end{eqnarray}
For a given quantum Hamiltonian of conduction electron subsystem, computing $\mathbf{S}_\mathrm{CD}^i(t)$ microscopically  generates {\em all} relevant spin torque terms $\propto \mathbf{S}_\mathrm{CD}^i \times \mathbf{M}_i$ in the LLG equation for $\mathbf{M}_i(t)$.  In Eq.~\eqref{eq:spin}, ${\bm \sigma}=(\hat{\sigma}_x,\hat{\sigma}_y,\hat{\sigma}_z)$ is the vector of the Pauli matrices, and one has to subtract~\cite{Nikolic2018,Mahfouzi2013} any nonzero equilibrium spin density (present in the absence of current) by using the NEGF expression for the equilibrium density matrix~\cite{Stefanucci2013},  
\begin{equation}
{\bm \rho}_\mathrm{eq} = -\frac{1}{\pi} \int\limits_{-\infty}^{+\infty} dE\, \mathrm{Im}\, \mathbf{G}(E) f(E), 
\end{equation}
where  $\mathbf{G}(E)$ is the retarded Green function (GF) in equilibrium and $f(E)$ is the Fermi distribution function (identical for both reservoirs in equilibrium)

\begin{figure}
	\includegraphics[scale=1.0,angle=0]{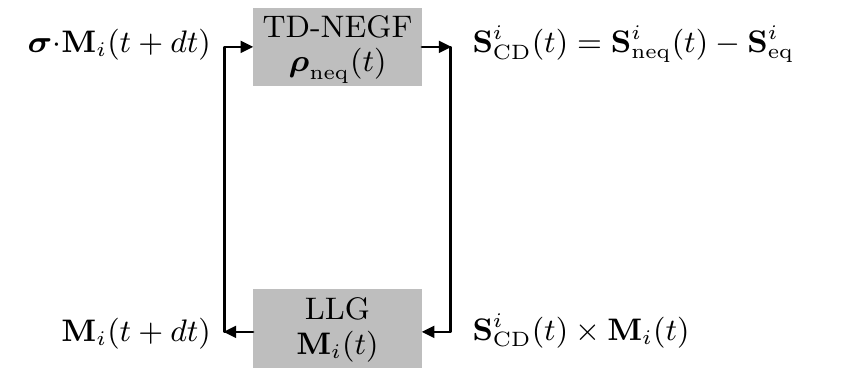}
	\caption{Scheme of TD-NEGF+LLG self-consistent loop in which TD-NEGF calculations supply current-driven part of electronic nonequilibrium spin density 
	$\mathbf{S}_\mathrm{CD}(t)$ defined in Eq.~\eqref{eq:spin}. This quantity determines spin torque entering into the LLG equation for the dynamics of classical vectors 
	$\mathbf{M}_i$ representing localized magnetic moments. In turn, the LLG equations supplies the time-dependent \mbox{$s$-$d$} interaction term, ${\bm \sigma} \cdot \mathbf{M}_i(t)$, for the quantum Hamiltonian of the conduction electrons.}
	\label{fig:blockdiagram}
\end{figure}
\begin{figure*}
	\includegraphics[scale=1.5,angle=0]{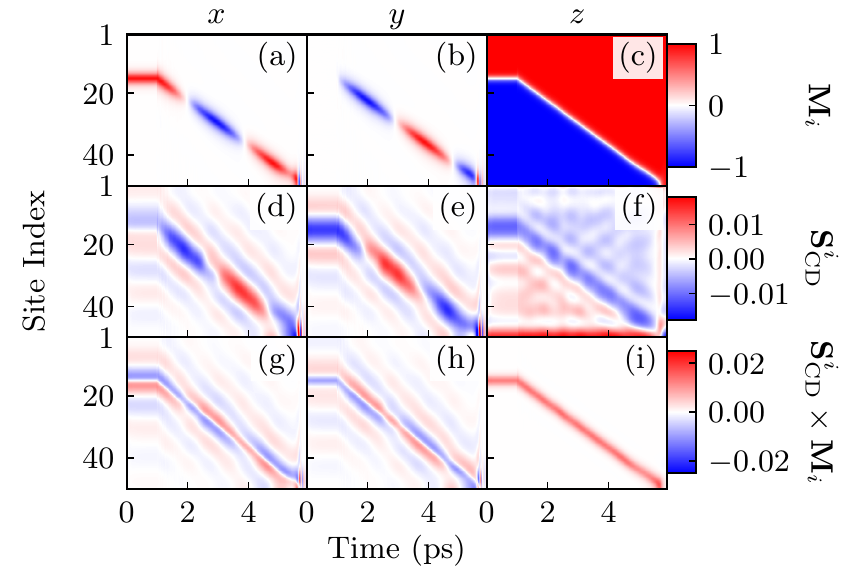}
	\caption{Spatio-temporal profiles of the components of: (a)--(c) localized magnetic moments $\mathbf{M}_i(t)$; (d)--(f) current-driven nonequilibrium spin  density  $\mathbf{S}_\mathrm{CD}^i(t)$ defined in Eq.~\eqref{eq:spin}; and (g)--(i) spin torque $\mathbf{T} \propto \mathbf{S}_\mathrm{CD}(t) \times \mathbf{M}_i(t)$ acting on the localized magnetic moments. The Fermi energy is $E_F^b=0.05$ eV, $s$-$d$ interaction between conduction electrons and  localized magnetic moments is \mbox{$J_{sd}=0.1$ eV} and the applied DC bias voltage is $eV_b=0.05$ eV. The  profiles are steady (after transient time following switching of DC bias voltage at $t=0$) for $t < 1$ ps, where DW is fixed at $X_\mathrm{DW}=15$, but they become time-dependent after coupling to LLG dynamics is turned on for $t \ge 1$ ps.}
	\label{fig:fig2}
\end{figure*}
\begin{figure}
	\includegraphics[scale=1.0,angle=0]{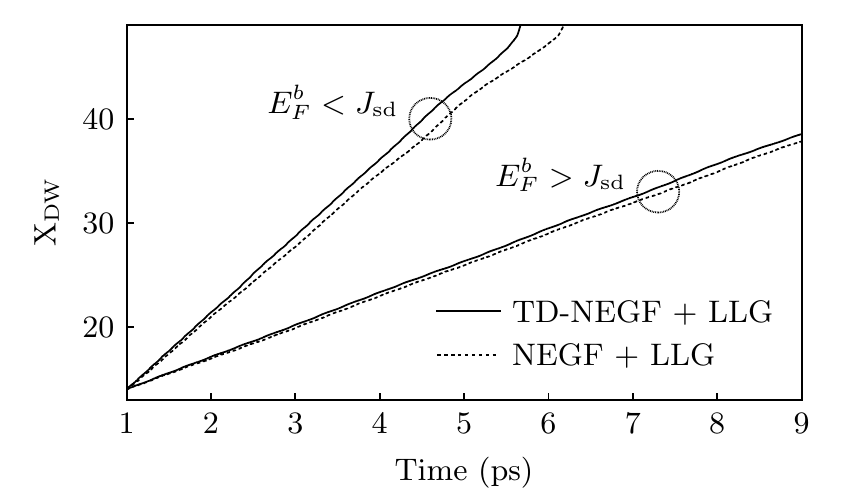}
	\caption{The position $X_\mathrm{DW}$ of the DW center as a function of time for $E_F^b=0.05$ eV $<$ $J_{sd}=0.1$ eV and $E_F^b=0.15$ eV  $>$ $J_{sd}=0.1$ eV computed using TD-NEGF+LLG (solid lines) and NEGF+LLG (dashed lines) formalisms applied to the device in Fig.~\ref{fig:fig1} with bias voltage $eV_b=0.05$ eV. The corresponding time evolution of localized magnetic moments $\mathbf{M}_i(t)$ is provided as a movie in the Supplemental Material~\cite{sm}.}
	\label{fig:fig3}
\end{figure}
\begin{figure*}
	\includegraphics[scale=1.0,angle=0]{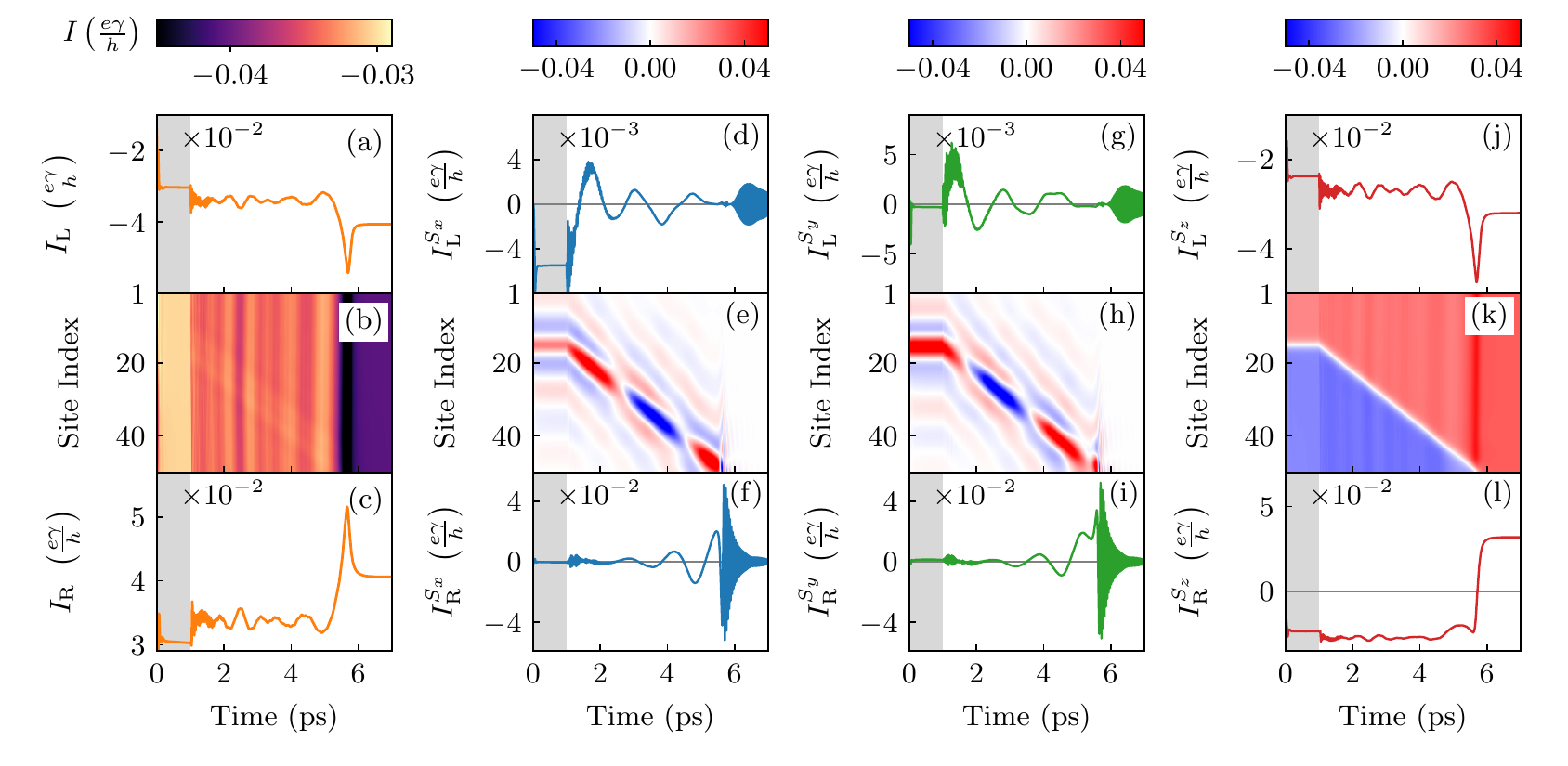}
	\caption{Charge current in the (a) left and (c) right NM leads, as well as spin currents in the (d),(g),(j) left and (f),(i),(l) right NM leads. These currents are steady (after transient time following switching of DC bias voltage at $t=0$) in the shaded area (for $t < 1$ ps), where DW is fixed at $X_\mathrm{DW}=15$, but they become time-dependent after coupling to LLG dynamics is turned on for $t \ge 1$ ps. Panels (b) and (e),(h),(k) depict spatio-temporal  profile of the local bond charge and spin currents, respectively. The parameters  are chosen as $E_F^b =0.05$ eV, $J_{sd} = 0.1$ eV and $eV_b=0.05$ eV.}
	\label{fig:fig4}
\end{figure*}
%

%The DW motion is governed by the interplay between driving force, which can be either externally applied magnetic field or injected current, and the complex energy %landscape generated by possible pinning centers along the ferromagnetic nanowire hosting the DW~\cite{Jiang2010}.

In this study, we develop a {\em numerically exact} [i.e., equivalent to summing all terms in the above mentioned power series expansion of ${\bm \rho}_\mathrm{neq}(t)$] approach denoted as TD-NEGF+LLG. As explained schematically in Fig.~\ref{fig:blockdiagram}, TD-NEGF+LLG employs ${\bm \rho}_\mathrm{neq}(t)$ in Eq.~\eqref{eq:timedm} to obtain $\mathbf{S}_\mathrm{CD}^i(t)$ via Eq.~\eqref{eq:spin}; which is then coupled to the LLG equation for $\mathbf{M}_i(t)$; which, in turn, is used to obtain ${\bm \rho}_\mathrm{neq}(t)$ at the next time step. The paper is organized as follows. The details of TD-NEGF+LLG framework are introduced in Sec.~\ref{sec:tdnegf}. To demonstrate richness of novel insights made possible by this framework, we apply it to widely studied~\cite{Tatara2008,Lee2004a,Stiles2007,Li2004,Li2004a,Thiaville2005,Thiaville2007,Martinez2009,Boone2010,Chureemart2011,Pivano2017} problem of current-driven DW motion along clean magnetic nanowire attached to two normal metal (NM) leads, where the injected unpolarized charge current from the NM leads is steady in Sec.~\ref{sec:steady} or pulsed  in Sec.~\ref{sec:pulse}. In Sec.~\ref{sec:spinmotive}, we employ a toy system of three precessing noncollinear spins to compare nonperturbative results from TD-NEGF for pumped charge current by this system to predictions  of a perturbative analytical formula of the so-called spin-motive force (SMF) theory~\cite{Barnes2007,Zhang2009b} for time-dependent magnetization textures, thereby delineating the limits of its validity. We conclude in Sec.~\ref{sec:conclude}.   

\section{TD-NEGF+LLG framework}\label{sec:tdnegf}

To make the discussion transparent, we use an example of a N\'{e}el DW illustrated in Fig.~\ref{fig:fig1} and described by a smooth function~\cite{Ohe2006a} of the position $x_i$ of site $i$ along the $x$-axis, 
\begin{equation}\label{eq:dwshape}
\mathbf{M}_i(t=0) = ([\cosh(X_\mathrm{DW}-x_i)/W]^{-1}, 0, \tanh(X_\mathrm{DW}-x_i)/W). 
\end{equation}
Its localized magnetic moments $\mathbf{M}_i$ lie entirely in the plane when the current is zero. Here $X_\mathrm{DW}$ is the coordinate of the DW center and $W=1a$ is its width (in the units of the lattice spacing $a$). The interaction between localized magnetic moments, whose direction at site $i$ is specified by unit vector $\mathbf{M}_i$ while their magnitude is $\mu_M$, is described by the {\em classical} Hamiltonian
\begin{eqnarray}\label{eq:heisenberg}
\mathcal{H} & = &  -J \sum_{\langle ij \rangle} \mathbf{M}_{i} \cdot \mathbf{M}_{j} -J_{sd} \sum_{i} \mathbf{S}^{i}_\mathrm{CD} \cdot 
\mathbf{M}_{i} \nonumber \\
\mbox{} && -K \sum_{i} (M_i^z)^{2} + D \sum_{i} (M_i^y)^{2}.
\end{eqnarray}
Besides Heisenberg term with exchange interaction between the nearest neighbors of strength \mbox{$J=0.1$ eV}, this includes \mbox{$s$-$d$} interaction between conduction electrons and  localized magnetic moments of strength \mbox{$J_{sd}=0.1$ eV}; magnetic anisotropy (along the $z$-axis) of strength \mbox{$K=0.025$ eV}; and demagnetization (along the $y$-axis) of strength \mbox{$D=0.029$ meV} (corresponding to the demagnetizing field of $\simeq 1$~T). The Hamiltonian in Eq.~\eqref{eq:heisenberg} determines the effective magnetic field acting on each localized  magnetic moment,  $\mathbf{B}_{\rm eff}^{i} = - \frac{1}{\mu_M} \partial \mathcal{H} /\partial \mathbf{M}_{i}$, which is inserted into the atomistic LLG equation (for simplicity, without noise term required at nonzero temperature)~\cite{Stiles2007,Evans2014,Chureemart2011}
\begin{equation}\label{eq:llg}
\frac{\partial\mathbf{M}_{i}}{\partial t} =
-\frac{g}{1 + {\lambda}^{2}}
\left[
\mathbf{M}_{i} \times \mathbf{B}_{\rm eff}^{i} 
+
\lambda \mathbf{M}_{i} \times 
\left(
\mathbf{M}_{i} \times \mathbf{B}_{\rm eff}^{i}
\right)
\right].
\end{equation}
Here  $g$ is the gyromagnetic ratio and the intrinsic Gilbert damping parameter~\cite{Gilmore2007} is chosen $\lambda=0.01$ as found in many realistic magnetic nanowires~\cite{Thiaville2005,Thiaville2007,Taniguchi2015}. Such coupled LLG equations are solved by the Heun numerical scheme~\cite{Evans2014}.

The conduction electron subsystem is described by the {\em quantum} Hamiltonian for a one-dimensional (1D) tight-binding (TB) model of a magnetic nanowire 
\begin{equation}\label{eq:tbh}
\hat{H}_\mathrm{TB} = - \gamma \sum_{\langle ij \rangle}  \hat{c}^\dagger_i \hat{c}_j  -  J_{sd} \sum_i  \hat{c}^\dagger_i {\bm \sigma} \cdot \mathbf{M}_i(t) \hat{c}_i,
\end{equation}
where $\hat{c}_i^\dagger=(\hat{c}_{i\uparrow}^\dagger \  \ \hat{c}_{i\downarrow}^\dagger)$ is a row vector containing operators $\hat{c}_{i\sigma}^\dagger$ which create an electron with spin $\sigma=\uparrow,\downarrow$ at site $i$; $\hat{c}_i$ is a column vector containing the corresponding annihilation operators; and $\gamma=1$ eV is the nearest-neighbor hopping. The TB chain described by Eq.~\eqref{eq:tbh}  is attached [Fig.~\ref{fig:fig1}] to two semi-infinite normal metal (NM) leads, modeled by the same Hamiltonian in Eq.~\eqref{eq:tbh} but with $J_{sd}=0$. We inject through the NM leads conventional unpolarized charge current using either DC bias voltage $V_b$, applied as electrochemical potential difference, $\mu_\mathrm{L}=E_F+eV_b/2$ and $\mu_\mathrm{R}=E_F-eV_b/2$  between the macroscopic reservoirs into which the left (L) and right (R) leads are assumed to terminate, or 
voltage pulses of different shape (see Fig.~\ref{fig:fig6} for illustration) whose amplitude is the same as the DC bias voltage. We quote the Fermi energy $E_F^b=E_F-E_b$ with respect to the bottom of the band $E_b=-2.0\gamma$ of the NM leads.

The Hamiltonian in Eq.~\eqref{eq:tbh} contains time-dependent term due to $\mathbf{M}_i(t)$ supplied [Fig.~\ref{fig:blockdiagram}] by solving the system of LLG equations displayed as  Eq.~\eqref{eq:llg}. Thus, rigorously it must be treated by some approach of time-dependent nonequilibrium quantum statistical mechanics which can yield ${\bm \rho}_\mathrm{neq}(t)$ in Fig.~\ref{fig:blockdiagram}. The TD-NEGF formalism~\cite{Stefanucci2013,Gaury2014} offers a route to ${\bm \rho}_\mathrm{neq}(t)$, as shown in Eq.~\eqref{eq:timedm}. It operates with two fundamental quantities~\cite{Stefanucci2013}---the retarded \mbox{$G^{\sigma \sigma^\prime}_{ii'}(t,t')=-i \Theta(t-t') \langle \{\hat{c}_{i\sigma}(t) , \hat{c}^\dagger_{i'\sigma'}(t')\}\rangle$} and the lesser \mbox{$G^{<,\sigma\sigma'}_{ii'}(t,t')=i \langle \hat{c}^\dagger_{i'\sigma'}(t') \hat{c}_{i\sigma}(t)\rangle$} GFs which describe the density of available quantum states and how electrons occupy those states, respectively.  For the device in Fig.~\ref{fig:fig1} we solve a matrix integro-differential equation~\cite{Croy2009,Popescu2016}
\begin{equation}\label{eq:master}
i\hbar \frac{d {\bm \rho}_\mathrm{neq}}{dt} = [\mathbf{H}_\mathrm{TB}, {\bm \rho}_\mathrm{neq}] + i \sum_{p=\mathrm{L,R}} [{\bm \Pi}_p(t) + {\bm \Pi}_p^\dagger(t)],
\end{equation}
which can be viewed as the exact master equation for an open finite-size quantum system, described by $\mathbf{H}_\mathrm{TB}$, which is attached via semi-infinite leads to much larger macroscopic reservoirs. We use convention in which bold-face symbols denote matrices in orbital$\otimes$spin vector space, where the size of the orbital space is equal to the number of sites ($i=1$--50 is chosen for the central magnetic nanowire region in Fig.~\ref{fig:fig1}) and the size of the spin space is two. The matrix 
\begin{equation}\label{eq:current}
{\bm \Pi}_p(t) = \int_{t_0}^t dt_2\, [\mathbf{G}^>(t,t_2){\bm \Sigma}_p^<(t_2,t) - \mathbf{G}^<(t,t_2){\bm \Sigma}_p^>(t_2,t) ],
\end{equation}
is expressed in terms of the lesser/greater GF and the corresponding lesser/greater self-energies ${\bm \Sigma}_p^{>,<}(t_2,t)$~\cite{Stefanucci2013},  whose numerical construction in order to convert Eq.~\eqref{eq:master} into a system of ordinary differential equations can be found in Ref.~\cite{Popescu2016}.  Equation~\eqref{eq:current} yields charge current in lead $p=\mathrm{L,R}$ of the device, 
\begin{equation}\label{eq:leadcharge}
I_p(t)=\frac{e}{\hbar} \mathrm{Tr}\, [{\bm \Pi}_p(t)], 
\end{equation}
as well as spin currents 
\begin{equation}\label{eq:leadspin}
I_p^{S_\alpha}(t)=\frac{e}{\hbar} \mathrm{Tr}\, [\hat{\sigma}_\alpha {\bm \Pi}_p(t)].
\end{equation}
We use the same units for both types of current---$I_p=I_p^\uparrow + I_p^\downarrow$ and $I_p^{S_\alpha}=I_p^\uparrow - I_p^\downarrow$---defined in terms of spin-resolved charge currents $I_p^\sigma$ for $\sigma=\uparrow,\downarrow$ along the $\alpha$-axis. The local (bond) charge  current~\cite{Nikolic2006} between sites $i$ and $j$ is computed as 
\begin{equation}\label{eq:bondcharge}
I_{i \rightarrow j}(t) =\frac{e\gamma}{i\hbar} \mathrm{Tr}_\mathrm{spin} \, \left[\rho_\mathrm{neq}^{ij}(t) - \rho_\mathrm{neq}^{ji}(t) \right], 
\end{equation}
and the local spin currents are given by 
\begin{equation}\label{eq:bondspin}
I_{i \rightarrow j}(t) = \frac{e\gamma}{i\hbar} \mathrm{Tr}_\mathrm{spin} \, \left[ \hat{\sigma}_\alpha \left\{ \rho_\mathrm{neq}^{ij}(t) - \rho_\mathrm{neq}^{ji}(t) \right\} \right]. 
\end{equation}
The computational complexity of  TD-NEGF calculations stems from the memory effect---the entire history must be stored in order to accurately evolve the NEGFs. For efficient calculation over long times and for large number of simulated sites, we employ newly developed TD-NEGF algorithms~\cite{Croy2009,Popescu2016} which scale linearly~\cite{Gaury2014} in the number of time steps.

We also compare our TD-NEGF+LLG framework to related efforts toward hybrid time-dependent-quantum/time-dependent-classical description of systems where conduction electrons interact with classical localized magnetic moments. Such an approach introduced in Ref.~\cite{Sayad2015} has the same feedback loop illustrated in Fig.~\ref{fig:blockdiagram}, but it considers electronic subsystems as a closed quantum system (e.g., as described by finite length TB chain~\cite{Sayad2015}) whose master equation in Eq.~\eqref{eq:master}, therefore, does not contain second term on the right-hand side. This makes it unsuitable for modeling of spintronic devices where one has to inject or collect spin and charge current through the attached semi-infinite leads. They also play an essential role by converting discrete spectrum of the central region into a continuous one, which ensures that current reaches steady-state in the long time limit after DC bias voltage is applied, even without explicit modeling of inelastic scattering processes. The approach of Ref.~\cite{Hammar2016} does include semi-infinite leads and macroscopic reservoirs into which they terminate, but it executes variety of approximations to make possible analytical solution for junctions containing a single classical localized spin, so it is not suitable for spatially extended spintronic devices with many coupled classical spins which require numerical modeling. The quantum part of both approaches~\cite{Sayad2015,Hammar2016} generates effectively a non-Markovian LLG equation due to additional time-retarded damping, on the top of intrinsic Gilbert damping (arising from combined effects of spin-orbit coupling and electron-phonon interaction~\cite{Gilmore2007}) that we take into account by using nonzero $\lambda$ in Eq.~\eqref{eq:llg}. Our TD-NEGF+LLG framework also contains time-retarded damping whose memory kernel properties will be discussed in future studies.

\section{DW motion driven by steady current: Spin and charge pumping}\label{sec:steady}

Evolving  ${\bm \rho}_\mathrm{neq}(t)$ via Eq.~\eqref{eq:master} requires time step \mbox{$\delta t=0.1$ fs} for numerical stability. The spatio-temporal profile of $\mathbf{S}_\mathrm{CD}^i(t)$ shown in Fig.~\ref{fig:fig2}(d)--(f) is obtained by plugging in thus evolved ${\bm \rho}_\mathrm{neq}(t)$ into Eq.~\eqref{eq:spin}. This is supplied to a system of LLG equations for $\mathbf{M}_i(t)$, whose spatio-temporal profile is shown in Fig.~\ref{fig:fig2}(a)--(c), where we use the same time step \mbox{$\delta t=0.1$ fs}. The noncollinearity at a given time between $\mathbf{S}_\mathrm{CD}^i$ [Fig.~\ref{fig:fig2}(d)--(f)] and $\mathbf{M}_i$ [Fig.~\ref{fig:fig2}(a)--(c)] generates spin torque $\mathbf{T} \propto \mathbf{S}_\mathrm{CD}^i \times \mathbf{M}_i$ on the DW ($T_x$ and $T_z$ determine damping-like torque and $T_y$ determines field-like torque~\cite{Ralph2008,Nikolic2018}) whose spatio-temporal profile is shown in Fig.~\ref{fig:fig2}(g)--(i). Figure~\ref{fig:fig2}(b), as well as movies showing complete time evolution of $\mathbf{M}_i(t)$ in the Supplemental Material~\cite{sm}, demonstrate how current-induced spin torque distorts moving DW with respect to the equilibrium N\'{e}el configuration by generating nonzero $M_i^y \neq 0$ component. 

Since TD-NEGF also captures transient charge and spin currents after the DC bias voltage is turned on at \mbox{$t=0$}, we first evolve conduction electron subsystem (during $t < 1$ ps in Fig.~\ref{fig:fig2}) with fixed DW (i.e.,  without coupling to LLG equations) until such currents become  steady.  This ensures that at $t=1$ ps, when LLG dynamics is turned on, spatial profile of ${\bf S}_\mathrm{CD}^i(t)$ computed by TD-NEGF and NEGF formalisms are identical. The position of the DW center as a function of time in Fig.~\ref{fig:fig3} computed by NEGF+LLG is similar to LLG result obtained in Fig.~1 of Ref.~\cite{Stiles2007}. On the other hand, it differs from LLG results of Ref.~\cite{Li2004,Li2004a} and related NEGF+LLG results of Ref.~\cite{Ohe2006a} where $X_\mathrm{DW}$ becomes saturated after relatively short time (i.e., DW motion comes quickly to a halt) for $E_F<J_{sd}$,  while DW continues to move  for $E_F>J_{sd}$ with $X_\mathrm{DW}$ exhibiting high-frequency oscillations (i.e., regularly accelerating and slowing down of the DW) due to the excitation of the spin waves~\cite{Li2004a,Ohe2006a}. This discrepancy could be due to time-retarded damping~\cite{Sayad2015,Hammar2016}, present in TD-NEGF+LLG but absent in NEGF+LLG framework, which can affect strongly~\cite{Bose2011} spin-wave excitation. Most importantly, TD-NEGF+LLG framework predicts faster DW motion in Fig.~\ref{fig:fig3} when compared to our NEGF+LLG results. This can be explained by additional torque exerted onto the DW by the  pumped~\cite{Tserkovnyak2005,Chen2009} spin currents of electrons in the presence of localized magnetic moment precession, as depicted in the movies in the Supplemental Material~\cite{sm}, which is a purely time-dependent quantum-mechanical effect {\em absent} in either LLG or NEGF+LLG simulations. Although the difference between the TD-NEGF+LLG and NEGF+LLG  results in Fig.~\ref{fig:fig3} is small over \mbox{$\sim 10$ ps} time interval considered, a much larger one can be extrapolated as one approaches \mbox{$\sim 10$ ns} typical time of DW motion in experiments and applications~\cite{Thomas2006,Thomas2007,Thomas2010,Chauleau2010,Taniguchi2015}. 

The TD-NEGF+LLG framework allows us to obtain explicitly time-dependent charge $I_p$ [Fig.~\ref{fig:fig4}(a),(c)] and spin $I_p^{S_\alpha}$ [Fig.~\ref{fig:fig4}(d),(g),(j),(f),(i),(l)] currents flowing into the NM leads in the course of DW motion, as well as spatio-temporal profiles of bond charge $I_{i \rightarrow j}$ [Fig.~\ref{fig:fig4}(b)] and bond spin  $I_{i \rightarrow j}^{S_\alpha}$ [Fig.~\ref{fig:fig4}(e),(h),(k)]  currents flowing between the nearest-neighbor sites. Note that these {\em time-dependent} currents are superimposed on the background of injected DC charge current, or DC spin current generated by spin-polarizing effect of the localized magnetic moments on injected DC current (the background values can be read from the flat lines within \mbox{$t < 1$ ps} interval in Fig.~\ref{fig:fig4}). 

\begin{figure}
	\includegraphics[scale=1.0,angle=0]{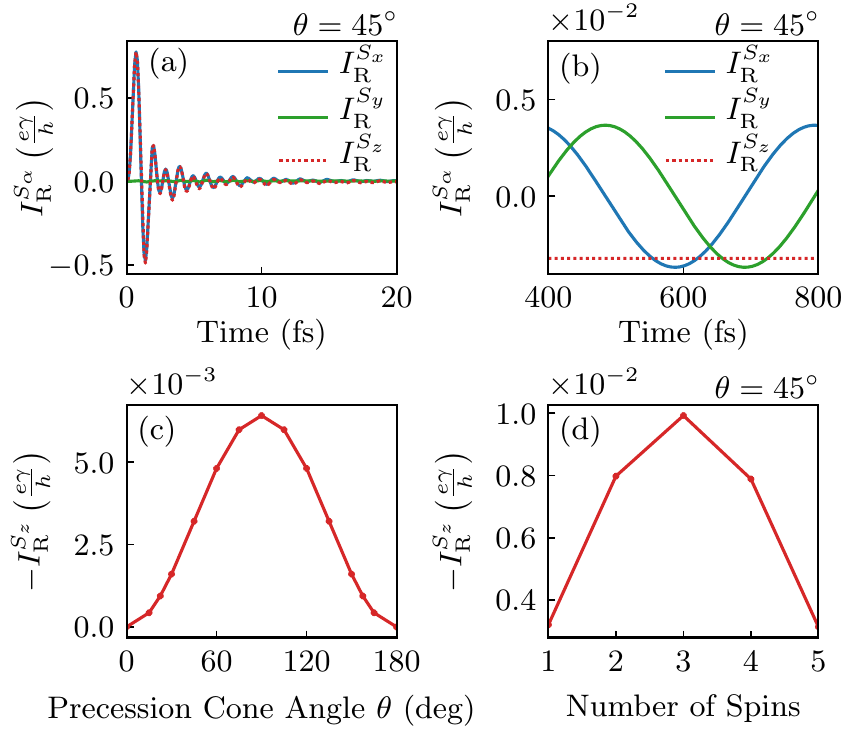}
	\caption{(a),(b) Time-dependence of spin currents $I^{S_\alpha}_\mathrm{R}(t)$ in the right lead of an infinite TB chain hosting a single magnetic moment in the middle precessing steadily with frequency \mbox{$\hbar \omega=1 \times 10^{-2}$ eV} and cone angle $\theta=45^\circ$ in the absence of any DC bias voltage (for illustration of this setup see Fig.~1 in Ref.~\cite{Chen2009}). Spin currents in the left lead have the same magnitude, but opposite direction. (c) $I^{S_z}_\mathrm{R}$, which is steady in panel (b), as a function of precession cone angle $\theta$. (d) Scaling of $I^{S_z}_\mathrm{R}$ with the number of precessing magnetic moments. The parameters are chosen as \mbox{$E_F^b=2$ eV} and $J_{sd}=1$ eV, as well as $J \equiv 0$ in the case of more than one precessing magnetic moment in panel (d).}
	\label{fig:fig5}
\end{figure}

Since movies of time evolution of $\mathbf{M}_i(t)$ in the Supplemental Material~\cite{sm} show that three localized magnetic moments around the propagating DW center are precessing, to gain intuition about how they induce spin and charge pumping in Fig.~\ref{fig:fig4} we first examine the simplest example of a single [Fig.~\ref{fig:fig5}(a)--(c)] or up to five [Fig.~\ref{fig:fig5}(d) magnetic moments $\mathbf{M}_i(t)$ precessing steadily with frequency $\omega$ and precession cone angle $\theta$ while being coupled to an infinite 1D TB chain~\cite{Chen2009}. Such setup---precessing spins in the center of 1D TB chain (for illustration see Fig.~1 in Ref.~\cite{Chen2009})---pumps {\em only} spin currents in both directions, as shown in Fig.~\ref{fig:fig5}(b). This problem is exactly solvable in the rotating frame, where our result in Fig.~\ref{fig:fig5}(b), after transient currents in Fig.~\ref{fig:fig5}(a) die away,  matches analytical formula derived in Ref.~\cite{Chen2009}, thereby also validating the accuracy of TD-NEGF numerical calculations. In particular, time-independent $I_p^{S_z}$ in Fig.~\ref{fig:fig5}(c) exhibits standard $\propto \sin^2 \theta$ dependence~\cite{Tserkovnyak2005} on the precession cone angle $\theta$. The maximum output in Fig.~\ref{fig:fig4}(d) is achieved by using three magnetic moments precessing together, which signifies interfacial nature~\cite{Tserkovnyak2005,Chen2009} of spin pumping. 

In addition, even single precessing magnetic moment can pump charge current with nonzero DC component on the proviso that the key requirement in the theory of quantum charge pumping by a time-dependent fields is satisfied~\cite{Vavilov2001,Moskalets2002,FoaTorres2005,Bajpai2018}---{\em breaking of left-right symmetry}. This requires to break inversion symmetry and/or time-reversal symmetry.  If both inversion and time-reversal symmetries are broken dynamically, the DC component of pumped charge current is $\propto \Omega$ at low frequencies, as found in standard example of quantum dot attached to two leads and exposed to two spatially separated potentials oscillating out-of-phase~\cite{Vavilov2001,Moskalets2002}. If only one of those two symmetries is broken, and this does not have to occur dynamically, the DC component of the pumped current  $\propto \Omega^2$ at low frequencies, as found in the case of single precessing magnetic moment with static potential barrier (breaking inversion symmetry) introduced into the 1D TB chain hosting the magnetic moment~\cite{Chen2009}. 

Armed with this intuition, we can interpret currents in Fig.~\ref{fig:fig4}(b),(e),(h),(k) as the consequence of moving DW center pumping spin and charge currents due to the dynamics of magnetic moments around the DW center depicted in the movies in the Supplemental Material~\cite{sm}. The pumped charge current arises because the DW itself breaks the left-right symmetry while localized magnetic moments around its center are driven into precession by spin torque, as visualized in Fig.~\ref{fig:fig2}. The collision of the DW with the right NM lead results in its annihilation, i.e., all $\mathbf{M}_i$ eventually point along the $z$-axis, which generates spike in the charge and spin currents in Fig.~\ref{fig:fig4} around $t \simeq 6$ ps. 

While variety of techniques have been developed to determined the position of a moving DW~\cite{Kim2017a,Singh2010,Krzysteczko2017}, they often have limitations in resolution or acquisition speed~\cite{Krzysteczko2017}. Figure~\ref{fig:fig4} shows that temporal profiles of $I_p^{S_x}(t)$ and $I_p^{S_y}(t)$ are tightly correlated with the DW position, as well as that their amplitude increases (decreases) as the DW approaches (distances from) the NM lead. Thus, converting these spin currents into AC voltage via the inverse spin Hall effect, which can be done experimentally with high efficiency~\cite{Wei2014}, offers an electrical measurement that precisely tracks the position of a single DW propagating along magnetic nanowire.

\section{TD-NEGF+LLG vs. spin-motive force theory for charge current pumped by time-dependent magnetic textures}\label{sec:spinmotive}

The spin-motive force (SMF)~\cite{Barnes2007,Zhang2009b,Kim2012b,Berger1986,Volovik1987,Stern1992,Saslow2007,Duine2008,Tserkovnyak2008a,Liu2011e,Lucassen2011,Yamane2011,Hals2015}  refers to pumping of charge and spin currents, or generation of voltage associated with pumped charge current, by {\em time-dependent noncoplanar and noncollinear} magnetic textures within conducting ferromagnets. In contrast to conventional electromotive force induced by change of magnetic flux through a circuit in accord with Faraday law of classical electromagnetism, SMF originates from spin and can appear even in static uniform external magnetic field (as long as such field generates dynamics of localized magnetic moments). The SMF has been invoked to explain experimental detection of electric voltage due the motion of magnetic DW~\cite{Yang2009}, magnetization reversal of nano-particles embedded in a MTJ~\cite{Hai2009} and gyration of magnetic vortex core~\cite{Tanabe2012}. Thus, SMF phenomenon is certainly related to charge current pumping explored in Sec.~\ref{sec:steady}, and in this Section we investigate this relationship in detail.

\begin{figure}
	\includegraphics[scale=1.0,angle=0]{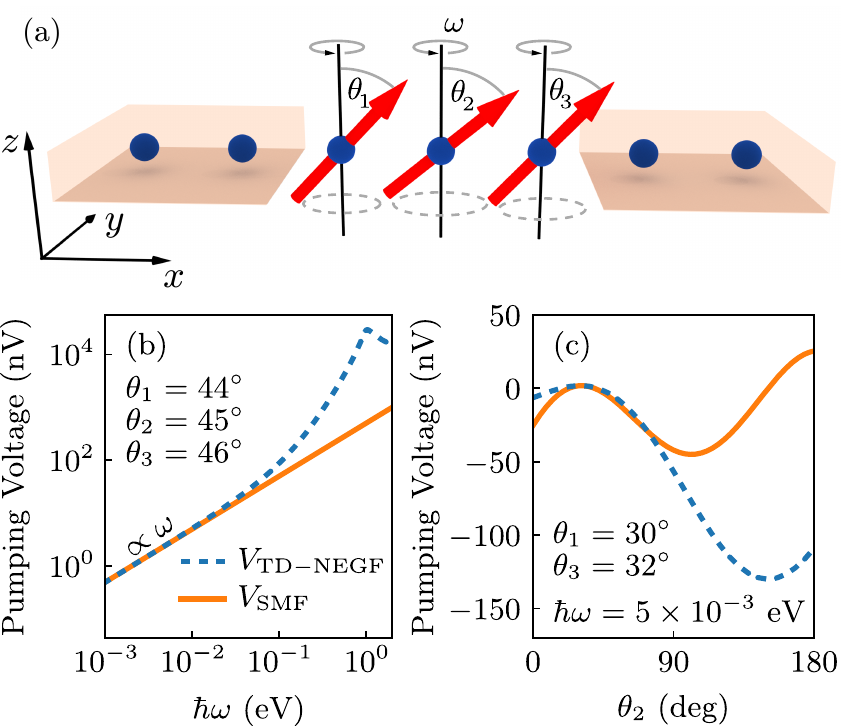}
	\caption{(a) Schematic view of a toy noncollinear and noncoplanar system, consisting of three localized magnetic moments in the middle of an infinite TB chain precessing at the same frequency $\omega$ but with different cone angles, which pumps spin and charge current into the semi-infinite leads in the absence of any bias voltage ($V_b \equiv 0$). The magnetic moments do not interact with each other ($J \equiv 0$), their $s$-$d$ interaction with electrons within the TB chain is \mbox{$J_{sd}=0.1$ eV} and electronic Fermi energy is chosen as $E_F^b=3.0$ eV. (b) The dc pumping voltage between the leads as a function of $\omega$ for small noncollinearity of magnetic moments, $\theta_1=44^\circ$, $\theta_2=45^\circ$ and $\theta_3=46^\circ$. (c) The dc pumping voltage between the leads as a function of $\theta_2$ while $\theta_1=30^\circ$ and $\theta_3=32^\circ$ are fixed, and all three localized magnetic moments are precessing at fixed frequency $\hbar \omega =5 \times 10^{-3}$ eV. The voltages in panels (b) and (c) are computed either from Eqs.~\eqref{eq:smfvoltage} and ~\eqref{eq:smfcharge} of the SMF theory~\cite{Barnes2007,Zhang2009b} (solid line) or numerically exactly from TD-NEGF formalism (dashed line).}
	\label{fig:fig8}
\end{figure}

The voltage associated with SMF between the edges of the wire lying along the $x$-axis~\cite{Barnes2007}
\begin{equation}\label{eq:smfvoltage}
V_\mathrm{SMF}=\frac{1}{G_0} \int j_x dx 
\end{equation}
is obtained from pumped local charge current~\cite{Zhang2009b} 
\begin{equation}\label{eq:smfcharge}
j_\alpha (\mathbf{r}) = \frac{P \sigma_0 \hbar}{2e} [\partial_t \mathbf{m}(\mathbf{r},t) \times \partial_\alpha \mathbf{m}(\mathbf{r},t)] \cdot \mathbf{m}(\mathbf{r},t),
\end{equation}
where $\partial_t = \partial/\partial t$ and $\partial_\alpha=\partial/\partial \alpha$ for $\alpha \in \{x,y,z \}$;  $\sigma_0=\sigma^\uparrow + \sigma^\downarrow$ is the total conductivity;  and \mbox{$P=(\sigma^\uparrow - \sigma^\downarrow)/(\sigma^\uparrow + \sigma^\downarrow)$} is the spin polarization of the ferromagnet. In general, conductivities $\sigma^\uparrow$ and $\sigma^\downarrow$ for the spin-$\uparrow$ and spin-$\downarrow$ bands depend on external magnetic field due to magnetoresistive effect, but this dependence can be neglected for transition metal ferromagnets. Similarly, part of the $3 \times 3$ tensor of pumped local spin current flowing along the $\alpha$-axis is given by the vector~\cite{Zhang2009b} 
\begin{equation}\label{eq:smfspin}
[j_\alpha^{S_x} (\mathbf{r}),j_\alpha^{S_y} (\mathbf{r}),j_\alpha^{S_z} (\mathbf{r})] = \frac{g \mu_B \hbar \sigma_0}{4e^2} [\partial_t \mathbf{m}(\mathbf{r},t) \times \partial_\alpha \mathbf{m}(\mathbf{r},t)]
\end{equation}
where $\mu_B$ is the Bohr magneton. 

Equations~\eqref{eq:smfcharge} and ~\eqref{eq:smfspin} contain only the lowest order time and spatial derivatives of magnetization~\cite{Freimuth2018}, so that comparing them to our nonperturbative results from TD-NEGF+LLG makes it possible to establish limits of validity of these equations. For this purpose, we employ a toy noncoplanar and noncollinear system consisting of three localized magnetic moments precessing with the same frequency $\omega$, which is illustrated in Fig.~\ref{fig:fig8}(a) and it is akin to a system analyzed in Fig.~\ref{fig:fig5}  but with different precession cone angles $\theta_1$, $\theta_2$ and $\theta_3$. Similarly to studies combining classical micromagnetics with SMF formula~\cite{Yamane2016}, the temporal dependence of these three localized magnetic moments is plugged into the discretized version of Eq.~\eqref{eq:smfcharge}
\begin{eqnarray}\label{eq:smfdiscrete}
j_x(i) & \propto & \frac{1}{a} [\partial_t \mathbf{M}_i(t) \times (\mathbf{M}_{i+1}(t)-\mathbf{M}_i(t))] \cdot \mathbf{M}_i(t)  \nonumber \\ 
 & \propto & \frac{1}{a}[\partial \mathbf{M}_i(t) \times \mathbf{M}_{i+1}(t)]\cdot \mathbf{M}_i(t),
\end{eqnarray} 
Since Eqs.~\eqref{eq:smfvoltage} and \eqref{eq:smfdiscrete} do not allow us to compute charge current flowing into the leads, we plug $j_x(i)$ from Eq.~\eqref{eq:smfdiscrete} into Eq.~\eqref{eq:smfvoltage} to obtain the SMF voltage $V_\mathrm{SMF}$ between the edges of the central region in Fig.~\ref{fig:fig8}(a). This is then compared to pumping voltage $V_\mathrm{TD-NEGF}=I_p/G$ in an open circuit computed using charge current $I_p$ in Eq.~\eqref{eq:leadcharge} pumped into NM leads and two-terminal conductance $G$ obtained from 
the Landauer formula.

For small noncollinearity between three magnetic moments in Fig.~\ref{fig:fig8}(a)---$\theta_1=44^\circ$, $\theta_2=45^\circ$ and $\theta_3=46^\circ$---voltages $V_\mathrm{SMF}$ and $V_\mathrm{TD-NEGF}$ track each other, while following $\propto \omega$ dependence, in Fig.~\ref{fig:fig8}(b) for all frequencies relevant for magnetization dynamics (the highest in the THz range, or \mbox{$\hbar \omega \sim 0.004$ eV}, are encountered in the dynamics of antiferromagnets~\cite{Jungfleisch2018}). However, if we fix the precession frequency and change angles between neighboring magnetic moments, we find increasing deviation  between $V_\mathrm{SMF}$ and $V_\mathrm{TD-NEGF}$ once the relative angles becomes $\gtrsim 10^\circ$ in Fig.~\ref{fig:fig8}(c), which can reach factor of two difference at large angles.

\section{DW motion driven by pulse current: Transient inertial displacement and spin and charge pumping}\label{sec:pulse}

\begin{figure}
	\includegraphics[scale=1.0,angle=0]{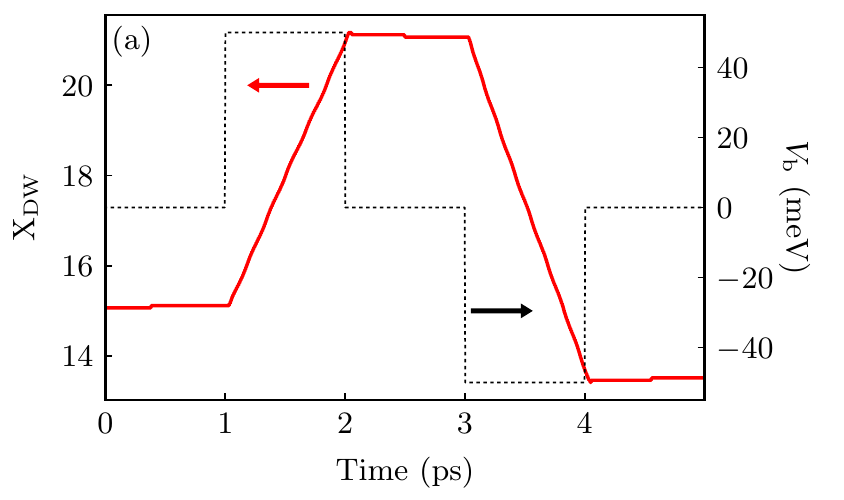}
	\includegraphics[scale=1.0,angle=0]{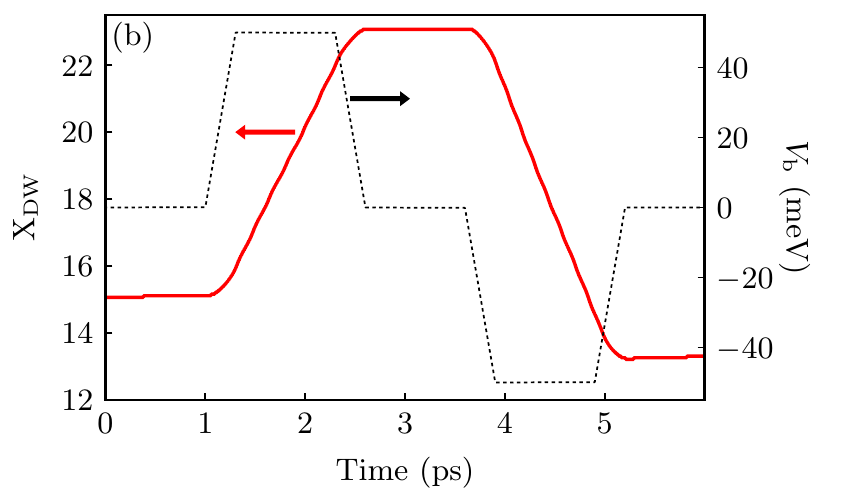}
	\caption{The position $X_\mathrm{DW}$ of the DW center as a function of time, where the DW motion is induced by applying: (a) a sequence of two successive rectangular  voltage pulses; or (b) a sequence of two successive trapezoidal voltage pulses. The temporal characteristics of the sequence of two pulses of opposite polarity is depicted by the dashed line, while their magnitude is the same as the DC bias voltage employed in Figs.~\ref{fig:fig3}, ~\ref{fig:fig4} and ~\ref{fig:fig5}. The parameters  are chosen as $E_F^b =0.05$ eV, $J_{sd} = 0.1$ eV and $eV_b=0.05$ eV.  The corresponding time evolution of localized magnetic moments $\mathbf{M}_i(t)$ is provided as two movies in the Supplemental Material~\cite{sm}.}
	\label{fig:fig6}
\end{figure}

The pulse-current-driven DM motion is of particular relevance for racetrack memory applications~\cite{Parkin2008,Parkin2015} where digital information is characterized by the orientation of the magnetic domain and data processing is carried out via current-induced DW motion. Thus, precise control of the position of the DW is required to achieve successful memory operation~\cite{Thiaville2007,Thomas2010,Chauleau2010,Taniguchi2015}. Although the DW displacement is related to current pulse duration, it is in general not linear relation due to {\em transient inertial displacement} (or automotion)~\cite{Thiaville2007,Thomas2010,Taniguchi2015,Pivano2017,Rhensius2010} appearing at current pulse onset and after pulse termination. Thus, too large transient inertial displacement will be detrimental for racetrack memory operation. The origin of transient inertial displacement is deformation of the DW leading to delayed response at the current onset and at the end of the current pulse, which then requires to tune the duration~\cite{Thiaville2007,Thomas2010,Chauleau2010,Taniguchi2015} and the shape (i.e., its rise and fall time)~\cite{Pivano2017} of the pulse. The experiments~\cite{Thomas2010,Chauleau2010,Taniguchi2015} and classical micromagnetic simulations~\cite{Thiaville2007,Chauleau2010,Pivano2017} typically employ short \mbox{$\sim$ ns} pulses, which generate higher DW velocities than longer \mbox{$\sim \mu$s} pulses due to easier depinning by an additional force on the DW during the pulse rise time or by a small mean distance between pinning centers.

\begin{figure*}
	\includegraphics[scale=1.0,angle=0]{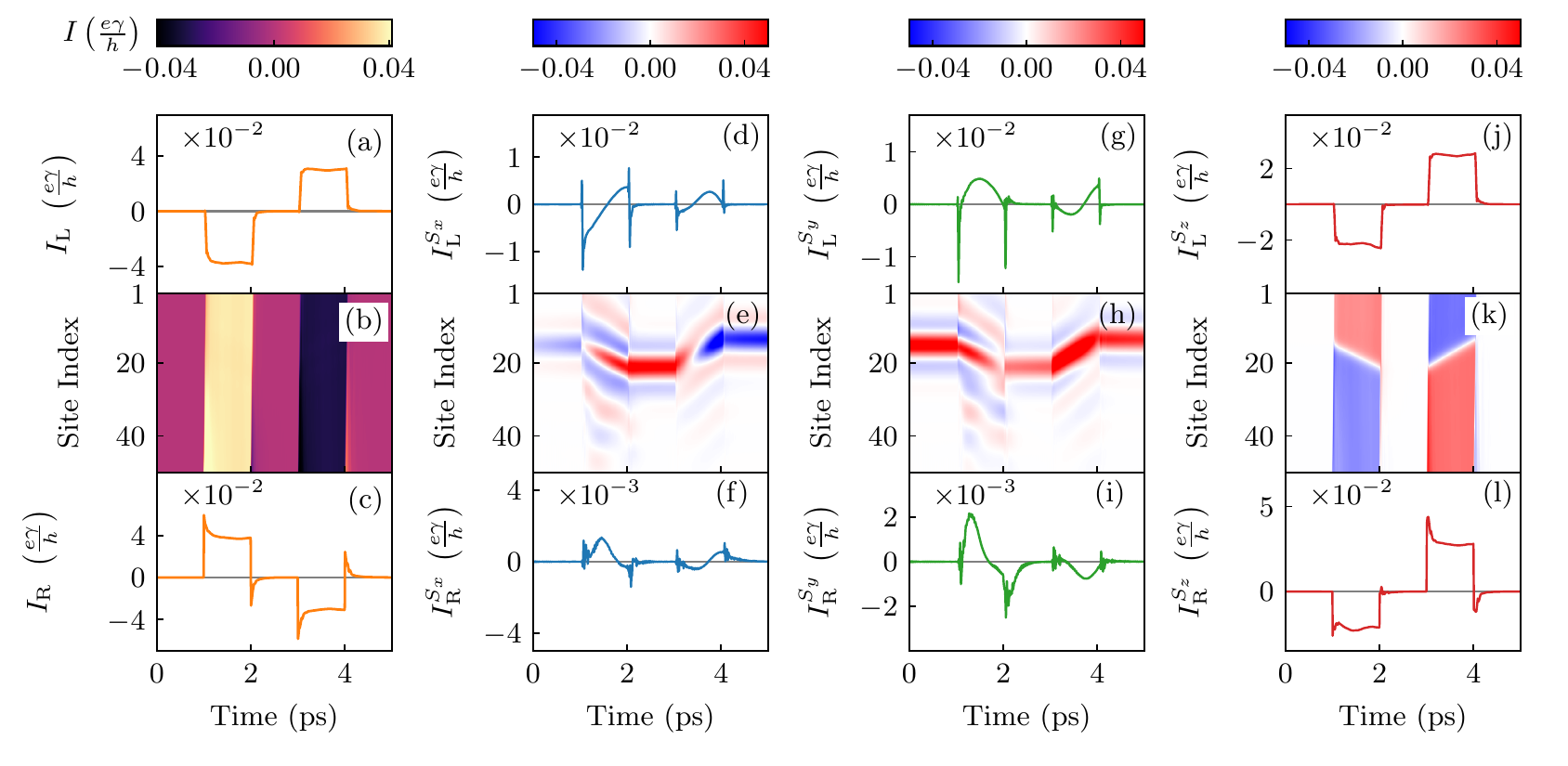}
	\caption{Time-dependence of charge current in the (a) left and (c) right NM leads, as well as spin currents in the (d),(g),(j) left and (f),(i),(l) right NM leads in the course of DW motion induced by a sequence of two successive rectangular voltage pulses depicted in Fig.~\ref{fig:fig6}(a). Panels (b) and (e),(h),(k) depict spatio-temporal  profile of the local bond charge and spin currents, respectively. The parameters  are chosen as $E_F^b =0.05$ eV, $J_{sd} = 0.1$ eV and $eV_b=0.05$ eV.}
	\label{fig:fig7}
\end{figure*}

We apply a sequence of two successive voltage pulses of opposite polarity whose temporal characteristics is shown in Fig.~\ref{fig:fig6} and whose magnitude is the same as the DC bias voltage used in Figs.~\ref{fig:fig3}, ~\ref{fig:fig4} and ~\ref{fig:fig5}.  We use rectangular [Fig.~\ref{fig:fig6}(a)] or trapezoidal [Fig.~\ref{fig:fig6}(b)] pulses of \mbox{$\sim$ ps} duration to understand basic physics and reduce computational expense.  The first pulse  drives DW forward (i.e., in the positive $x$-direction in Fig.~\ref{fig:fig1}) and the second pulse drives the DW backward, so that in the absence of transient inertial displacement the DW center would return to its initial position in Fig.~\ref{fig:fig6}. The transient inertial displacement in Fig.~\ref{fig:fig6}, $\delta X_\mathrm{DW} = X_\mathrm{DW}(t=0) - X_\mathrm{DW}(t=5 \ \mathrm{ps})$, is $\simeq 10$\% of the forward displacement generated by the first pulse and surprisingly close to  transient displacement observed in experiments where adiabatic STT drives the DW motion~\cite{Taniguchi2015}. On the other hand,  it is quite different from transient inertial displacement estimated~\cite{Thiaville2007} via simple formula, $\delta X_\mathrm{DW} = - W \delta \phi/\lambda$, using 1D model of the DW ($\delta \phi$ is the angle variation of the DW, which is $\delta \phi=\pi$ in the case of DW in Fig.~\ref{fig:fig1}). Thus, $\delta X_\mathrm{DW} \sim 10$--$100$ nm predicted by this formula, for typical Gilbert damping $\lambda \sim 0.01$--$0.1$ of magnetic nanowires, suggests transient displacement comparable or much larger than the bit size of the racetrack memory ($\sim 10$ nm bit size is required for racetrack memory to be competitive to other memory devices~\cite{Parkin2008,Parkin2015}) which would be a significant impediment for its operation. Since it contradicts experiments where much smaller $\delta X_\mathrm{DW}$ has been observed~\cite{Taniguchi2015}, classical micromagnetic simulations aiming to reproduce such observation have suggested~\cite{Taniguchi2015} that engineering of extrinsic pinning sites is necessary to obtain small $\delta X_\mathrm{DW}$. 

Conversely, small $\delta X_\mathrm{DW}$ we obtain in Fig.~\ref{fig:fig6} for perfectly clean nanowires suggests the importance of inclusion of time-dependent quantum transport effects, such as spin and charge pumping generated while the DW experiences acceleration and deceleration due to injected pulse current. Figure~\ref{fig:fig7} shows spin and charge currents in the NM leads, as well as locally between the sites of magnetic nanowire, which emerge upon applying a sequence of two rectangular pulses depicted in Fig.~\ref{fig:fig7}(a) and can be contrasted to the same information presented in Fig.~\ref{fig:fig5} for the case of applied DC charge current. The charge currents in Figs.~\ref{fig:fig7}(a) and ~\ref{fig:fig7}(c) do not follow the shape of the pulse due to additional charge current being pumped when the DW starts of stop moving. The same applies to $I_p^{S_z}$ spin current which in the absence of DW motion would quantify spin polarization $I_p^{S_z}/I$~\cite{Dolui2017} along the $z$-axis after injected unpolarized charge current becomes polarized via propagation through magnetic nanowire depicted in Fig.~\ref{fig:fig1}. The spikes in spin currents at the instants of time where the pulse rises or decays would introduce additional terms in LLG dynamics which are absent in classical micromagnetics.

\section{Conclusions}\label{sec:conclude}

In conclusion, we have developed a {\em multiscale} theoretical and computational framework which self-consistently couples {\em time-dependent nonequilibrium quantum 
statistical} description of conduction electrons with {\em time-dependent classical} description of localized magnetic moments. The TD-NEGF+LLG framework requires just time-dependent quantum 
and classical Hamiltonians, together with device geometry, as an input for computing the time evolution of the interacting electron--localized-magnetic-moments many-body system in {\em numerically exact} fashion. This can be contrasted with widely used classical micromagnetic simulations~\cite{Xiao2005,Berkov2008,Stiles2007,Evans2014,Baumgartner2017,Lee2004a,Stiles2007,Li2004,Li2004a,Thiaville2005,Thiaville2007,Martinez2009,Boone2010,Chureemart2011,Iwasaki2013a,Sampaio2013,Taniguchi2015}, where propagating conduction electrons appear only indirectly through phenomenological spin torque terms inserted by hand into the LLG equation; or with previous steady-state-NEGF+LLG attempts~\cite{Ohe2006a,Ohe2006b,Salahuddin2006,Xie2017,Ellis2017} to couple quantum electrons to classical 
localized magnetic moments where fast electrons are assumed to instantaneously respond to slow dynamics of localized magnetic moments so that noncommutativity of the electronic quantum Hamiltonian at different times is neglected. Using DW motion driven by steady or pulse injected charge current as an example, we essentially demonstrate introduction (via TD-NEGF) of 
quantum spin pumping by the dynamics of localized magnetic moments and additional time-retarded damping characterized by a memory kernel~\cite{Sayad2015,Hammar2016} into classical micromagnetics. In addition, we quantify {\em nonperturbatively} charge and spin currents pumped from  time-dependent magnetic texture into the attached NM leads. They can be used as signatures 
of the dynamics of DWs, skyrmions and spin superfluids~\cite{Kim2017b} that can be detected by standard charge transport measurements. The same problem of charge pumping by time-dependent magnetic textures is also tackled by the SMF theory~\cite{Barnes2007,Zhang2009b,Kim2012b,Berger1986,Volovik1987,Stern1992,Saslow2007,Duine2008,Tserkovnyak2008a,Liu2011e,Lucassen2011,Yamane2011,Hals2015,Yamane2016}. However,  its analytical formula in Eq.~\eqref{eq:smfcharge} is perturbative in nature (i.e., it contains only the lowest order time and spatial derivatives of magnetization), and direct comparison  with nonperturbative TD-NEGF+LLG framework shows [Fig.~\ref{fig:fig8}] that it fails when angles between neighboring localized magnetic moments exceed $\simeq 10^\circ$.

\begin{acknowledgments}
We thank L. E. F. Foa Torres for instructive discussions. M.~D.~P. and P.~P. were supported by ARO MURI Award No. W911NF-14-0247.  B. S. P., U. B. and B. K. N.  were supported by NSF Grant No. CHE 1566074. The supercomputing time was provided by  XSEDE, which is supported by NSF Grant No.  ACI-1548562.
\end{acknowledgments}
 
%BibTeX
%Windows:
%\bibliographystyle{C:/BIBTEX/prsty}
%\bibliography{C:/BIBTEX/qttg}

\end{document}